\documentclass[useAMS,usenatbib]{mn2e}

\usepackage{lscape}
\usepackage{amssymb}
\usepackage{longtable}
\usepackage{multicol}
\usepackage{graphicx}
\usepackage{url}

\title{An Accurate Cluster Selection Function for the J-PAS Narrow-Band wide-field survey.}
\author[B. Ascaso et al.]{B. Ascaso$^{1}$\thanks{E-mail:
begona.ascaso@obspm.fr},   
 N. Ben\'itez$^{2,3}$, R. Dupke$^{3,4,5}$, E. Cypriano$^{6}$, G. Lima-Neto$^{6}$, 
\newauthor C. L\'opez-Sanjuan$^{7}$,  J. Varela$^{7}$, J. S. Alcaniz$^{3}$,  T.~Broadhurst$^{8,9}$, A. J. Cenarro$^{7}$, 
\newauthor N. Chandrachani Devi$^{3}$,  L.~A.~D{\'{\i}}az-Garc{\'{\i}}a$^{7}$,  C. A. C. Fernandes$^{3}$,  
\newauthor C. Hern\'andez-Monteagudo$^{7}$,  S. Mei$^{1,10,11}$, C. Mendes de Oliveira$^{6}$, A. Molino$^{6}$, 
\newauthor  I. Oteo$^{12,13}$, W. Schoenell$^{2}$, L. Sodr\'e$^{6}$,  K. Viironen$^{7}$, A. Mar{\'{\i}}n-Franch$^{7}$\\
$^{1}$GEPI, Observatoire de Paris, PSL Research University,  CNRS, University of Paris Diderot, 61, Avenue de l'Observatoire 75014, Paris France \\
$^{2}$Instituto de Astrof\'isica de Andaluc\'ia (IAA-CSIC), Glorieta de la Astronom\'ia s/n, 18008, Granada, Spain\\
$^{3}$Observat\'orio Nacional-MCT, Rua Jos\'e Cristino, 77. CEP 20921-400, Rio de Janeiro-RJ, Brazil\\
$^{4}$Dept. of Astronomy, University of Michigan, Ann Arbor, MI 48109\\
$^{5}$Eureka Scientific, Oakland, California, 94602-3017\\
$^{6}$Instituto de Astronomia, Geof\'isica e Ci\^encias Atmosf\'ericas, Universidade de S\~ao Paulo, Cidade Universit\'aria, 05508-090, S\~ao Paulo, Brazil\\
$^{7}$Centro de Estudios de F\'isica del Cosmos de Arag\'on, Plaza San Juan 1, 44001 Teruel, Spain\\
$^{8}$Department of Theoretical Physics, University of the Basque Country UPV/EHU, 48080 Bilbao, Spain\\
$^{9}$IKERBASQUE, Basque Foundation for Science, Bilbao, Spain\\
$^{10}$University of Paris Denis Diderot, University of Paris Sorbonne Cit\'e (PSC), 75205 Paris Cedex  13, France\\
$^{11}$California Institute of Technology, Pasadena, CA 91125, USA\\
$^{12}$SUPA, Institute for Astronomy, University of Edinburgh, Royal Observatory, Blackford Hill, Edinburgh EH9 3HJ\\
$^{13}$European Southern Observatory, Karl-Schwarzschild-Str. 2, 85748 Garching, Germany\\}

\begin{document}

\date{Accepted . Received }


\maketitle

\label{firstpage}

\begin{abstract}

The impending Javalambre Physics of the accelerating universe Astrophysical Survey (J-PAS) will be the first wide-field survey of $\gtrsim$ 8500 deg$^2$ to reach the `stage IV' category. Because of the redshift resolution afforded by 54 narrow-band filters, J-PAS is particularly suitable for cluster detection in the range z$<$1. The photometric redshift dispersion is estimated to be only $\sim 0.003$ with few outliers $\lesssim$ 4\% for galaxies brighter than $i\sim23$ AB, because of the sensitivity of narrow band imaging to absorption and emission lines. Here we evaluate the cluster selection function for J-PAS using N-body+semi-analytical realistic mock catalogues. We optimally detect clusters from this simulation with the Bayesian Cluster Finder, and we assess the completeness and purity of cluster detection against the mock data. The minimum halo mass threshold we find for detections of galaxy clusters and groups with both $>$80\% completeness and purity is $M_h \sim 5 \times 10^{13}M_{\odot}$ up to $z\sim 0.7$. We also model the optical observable, $M^*_{\rm CL}$-halo mass relation, finding a non-evolution with redshift and main scatter of $\sigma_{M^*_{\rm CL} | M_{\rm h}}\sim 0.14 \,dex$ down to a factor two lower in mass than other planned broad-band stage IV surveys, at least. For the $M_{\rm h} \sim 1 \times 10^{14}M_{\odot}$ Planck mass limit, J-PAS will arrive up to $z\sim 0.85$ with a $\sigma_{M^*_{\rm CL} | M_{\rm h}}\sim 0.12 \, dex$. Therefore J-PAS will provide the largest sample of clusters and groups up to $z\sim 0.8$ with a mass calibration accuracy comparable to X-ray data.

\end{abstract}


\quad
\newpage
\section{Introduction}

We are living exciting times in Cosmology. Roughly 15 years after the discovery of the inconsistency of a $\Lambda=0$ universe with the magnitude-redshift observed relation for the Type Ia supernova \citep{riess98,perlmutter99}, many cosmological probes have pointed towards the universe passing through a phase of accelerated expansion: the cosmic microwave background anisotropies (e.g. \citealt{spergel03,komatsu11,planck11}), the Baryonic Acoustic Oscillations \citep{eisenstein05}, the clustering of galaxies (e.g. \citealt{reid10,sanchez12}) and the growth of massive galaxy clusters (e.g. \citealt{mantz10,rozo10,planck13}), among others.

One possible explanation for this acceleration can be postulated by introducing a new energy component in the form of a dark energy with negative pressure (for a review, see \citealt{mortonson13,weinberg13}). Consequently, the Dark Energy Task Force (DETF,  \citealt{albrecht06}) has been created, urging the cosmology community to invest its effort in understanding the origin and nature of the dark energy. 

With this goal in mind, a number of surveys have been planned for the upcoming years intending to constrain the values of the dark energy by a factor of $\ge$10 times better than at present (the so-called Stage IV surveys, \citealt{albrecht06}). Some of these surveys, cited in chronological order of predicted start off, are: the Javalambre Physics of the accelerating universe Astrophysical Survey (J-PAS \citealt{benitez14}), the Dark Energy Spectroscopic Instrument (DESI, \citealt{levi13}) survey, the Large Synoptic Sky Telescope (LSST, \citealt{ivezic08,lsst09}) survey, the Euclid \citep{laureijs11} survey and the Wide-Field Infrared Survey Telescope (WFIRST\footnote{http://www.ipac.caltech.edu/wfirst/}), among many others. For an excellent review of some of these surveys, we refer the reader to \cite{weinberg13}.

These surveys will follow complementary observational strategies, allowing constraints covering different regions of the cosmological parameter space for testing competing cosmological models.  \cite{ascaso15b}  recently explored the properties of the photometric redshift capabilities we may expect for the Euclid and LSST surveys, showing their different behaviours. In this work, we explore the unique photometric capability of the impending J-PAS multiple ($>$ 50) narrow-band survey now being commissioned. This survey samples the optical spectrum with 54 narrow-bands of  $\sim 145\AA$, providing photometric redshift accuracies close to what we would expect for a low resolution spectroscopic survey, where emission and absorption features will be detected photometrically. This kind of data, in the frontier between spectroscopic and photometric surveys, has never been explored before and it will allow us to reliably map the large-scale structure in 3D down to fainter magnitudes and larger areas than previous spectroscopic samples. To realise the full scientific potential of these data, new algorithms and techniques are being developed and tested and cosmological constraints will be forecasted \citep{xavier14,lopez-sanjuan14}.

In particular, we focus on the expected performance of J-PAS for galaxy clusters and groups related to cosmology. Clusters, by virtue of their extreme masses are of great importance for the purpose of setting cosmology constraints and the study of the large-scale structure (e.g. \citealt{mantz10,rozo10,planck13} or see for a review \citealt{allen11}).  Modeling accurately the cluster selection function and the uncertainties in the observable- theoretical mass relation is of major importance for extracting cosmological information from J-PAS. The amplitude of the cluster power-spectrum (e.g \citealt{lima05} and references herein) is expected to be rapidly evolving over the redshift range accessible to J-PAS (z$<$1.5), hence becoming very sensitive to the growth rate of structure (see review by \citealt{huterer15}).

Presently, only relatively weak constraints are derived from redshift space distortions (e.g. SDSS, Wiggle-Z, BOSS, etc) and even less derived from massive clusters owing to the current difficulties of completing even modest sized surveys with sufficient redshift information. Furthermore, current wide angle surveys sensitive to clusters through weak lensing (WL), X-ray measurements or the Sunyaev-Zel'dovich (SZ) effect are still dealing with alleviating the tension   between the different scaling of the SZ, WL and X-ray observable mass and the theoretical cluster mass in order to make the connection to the cosmological predictions (e.g. \citealt{vonderlinden14,rozo14b,planck15}).

Numerous techniques have been developed to detect galaxy clusters using X-ray data, the SZ effect, WL or optical/IR data (see \citealt{allen11} and references herein), being their selection function carefully modeled. For the optical and IR techniques, there is a large diversity of  selection functions depending on the technique or the survey considered (for a review see \citealt{ascaso13}).  For instance, only few surveys with large number of medium or narrow-bands (i.e. good photometric redshift resolution) had their cluster samples fully exploited (COSMOS, \citealt{bellagamba11}; ALHAMBRA, \citealt{ascaso15a}). Therefore, the selection function of clusters in narrow-band surveys such as J-PAS are still in the process of being explored.

In this paper we provide a comprehensive estimate of the cluster selection function for J-PAS, accounting for the expected photometric limits and redshift accuracy of our multi-narrow-bands. It must be stressed that J-PAS provides near optimal efficiency for separating cluster members from foreground and background galaxies because of its photometric redshift precision. This accuracy of the photometric redshifts is matched to the typical velocity dispersion of massive clusters, and therefore we can detect clusters above the noise to much lower masses and to higher redshifts than the wide-field surveys using conventional filters. This cluster selection function will be useful not only for providing cosmological forecasts from cluster counts but also for performing extended studies on galaxy evolution and large-scale structure in clusters

The structure of the paper is as follows. In Section \S2, we describe J-PAS, the survey used in this paper, giving an overview of its main characteristics. Section \S3 describes the simulation used in this work, the original mock catalogue and the posterior modification with  \texttt{PhotReal} intended to mimic the photometry and photometric redshifts realistically. In Section  \S4, we describe the photometric redshift properties of the J-PAS data and compare with other next-generation surveys such as the LSST and Euclid. Section \S5 presents the results of detecting galaxy clusters in these mocks. It first provides an explanation of the Bayesian Cluster Finder, the cluster detector used in this work. Then, it shows the results regarding the cluster selection function expected from the J-PAS-mock catalogue and it finally models the cluster observable-halo mass relation and its evolution with redshift. Finally,  we draw conclusions of the work in section \S6.

The cosmology adopted throughout this paper is $H_0=73$ ${\rm km s^{-1} Mpc^{-1}}$, $\Omega_M=0.25$, $\Omega_{\Lambda}=0.75$, $\Omega_K=0$, $\sigma_8=0.9$, corresponding to the cosmology assumed in the Millennium simulation used in this work for consistency. All the magnitudes in this work are provided in the AB system \citep{oke83}.

\section{The J-PAS survey}

The Javalambre-Physics of the Accelerating Universe Astrophysical Survey\footnote{http://j-pas.org/}   (J-PAS, \citealt{benitez14}) is the first  stage IV survey, starting in 2016. The observations will be taken from the Javalambre Survey Telescope (JST/T250), a new fully dedicated 2.5m telescope located at the Observatorio Astrof\'isico de Javalambre\footnote{http://oaj.cefca.es} in Teruel (Spain), using JPCam, a panoramic camera with a mosaic of 14 large-format CCDs amounting to 1200 Mpix, that provides an effective field of view of $\sim$4.7deg$^2$ (see \citealt{cenarro13,cenarro14,taylor14,marin-franch15}).

\begin{figure*}
\centering
\includegraphics[clip,angle=90,width=1.0\hsize]{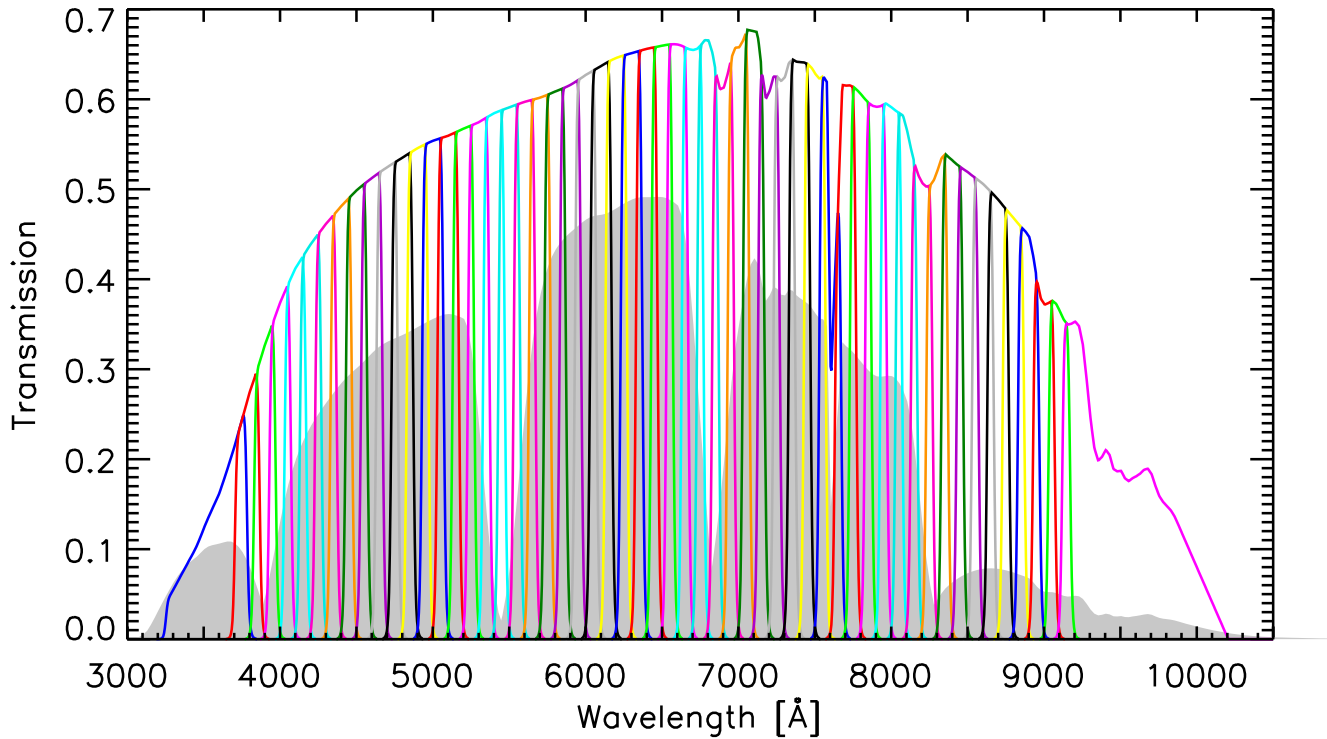} 
\caption{Transmission curves of the 54 narrow-band and 2 medium-band overlapping J-PAS filters spanning the optical range (color lines). The width of each narrow-band filter is $\sim 145$\AA\, and they are spaced by 100\AA.  For comparison the five SDSS filters are shown with gray shaded shape.}
\label{fig:transmissioncurves}
\end{figure*}

With the main purpose of constraining the dark energy parameters with at least 10 times higher precision than present surveys, J-PAS will image $\gtrsim$ 8500 $deg^2$ of the northern sky with 54 narrow-band filters plus 2 medium-band  and 3 broad-band $ugriz$-like filters in the whole optical range. Each narrow-band filter will have a width of $\sim 145$\AA\, and will be spaced by 100\AA. The filter transmission curves of the 54 narrow-band overlapping filters plus the two medium-band filters for J-PAS are displayed in Fig. \ref{fig:transmissioncurves} (see also \citealt{benitez14}). For comparison, we also plot the five broad-band filters of the Sloan Digital Sky Survey (SDSS). As we can see, the optical wavelength range for a low-redshift object will be sampled with more than 50 data points allowing, not only to recover a good estimation of the photometric redshift, but also to infer intrinsic properties of the galaxies.

The expected depth of the survey (5$\sigma$ detection magnitudes) for all the different bands are provided in Tables 3-5 in \cite{benitez14} from realistic simulations using the characteristics of the telescope, camera and site. In addition, we have created a synthetic $i$-band as a combination of the narrow-band filters of the survey, by following a similar procedure to that described in \cite{molino14,ascaso15a} for the ALHAMBRA survey. This has been made in order to use the same  pass-band to detect galaxy clusters as some other work in the literature (e.g. \citealt{postman02,olsen07,adami10,ascaso15a}).

Due to the large coverage of the visible spectrum, the expected photometric redshift accuracy will be $\Delta z\sim0.003(1+z)$ for more than $9\times10^7$ galaxies down to the flux limit of the survey \citep{benitez09a,benitez14}. This photometric redshift resolution makes this survey comparable to a low resolution integral field unity (IFU) of the northern sky.  

The excellent photometric redshift precision that J-PAS will achieve, makes this survey ideal for characterizing the overall galaxy population in terms of colours, morphology or chemical composition and therefore, for determining the cluster galaxy membership. 

\section{Simulating J-PAS}

In this paper, we use a mock catalogue generated by using the same procedure as in \cite{ascaso15b}. Indeed, we use the  500 $deg^2$ wide mock cone catalogue by \cite{merson13}\footnote{http://community.dur.ac.uk/a.i.merson/lightcones.html} designed to mimic Euclid and, we transform it into a J-PAS mock catalogue by using \texttt{PhotReal}. This technique, described in  \cite{ascaso15b}, obtains a new photometry and photometric error set for a particular survey to reproduce the observational properties of the galaxies with fidelity. After that, photometric redshifts have been derived by using BPZ2.0 (\citealt{benitez00}, Ben\'itez et al. in prep). In this section, we give a brief description of the mock catalogue construction.

\subsection{Light-cone original mock catalogue}

We use a mock catalogue constructed from the Millennium dark matter simulation  \citep{springel05}. The dark matter haloes have been populated with galaxies created through the semi-analytic galaxy formation model GALFORM  \citep{cole00,lagos11}. The light-cone is built from different simulation's snapshots up to $z=3$, allowing for interpolation between snapshots in order to properly model the evolution of structures along the line of sight. For a detailed explanation, we refer the reader to \cite{merson13}. 

The solid angle of the cone used is 500 $deg^2$, which is $\sim$ 16 times smaller than the actual surveyed area of J-PAS. While this fact does not affect the recovery of the photometric redshift accuracy; it might cause the absence of some rare, very massive clusters. However, these clusters are always well-identified through different techniques and for the purpose of characterizing the selection function, the results will remain virtually unchanged.

The mock catalogue contains a large quantity of physical parameters related with the dark matter haloes (namely: dark matter mass, dark matter ID, center of halo, galaxies belonging to each halo) and the galaxies (Galaxy ID, ra, dec, redshift, mass of cold gas, quiescent SFR in disk, stellar mass of galaxy, among others). The catalogue also includes 'spectroscopic' redshifts, defined as the cosmological redshift with peculiar velocities added.  Finally, the photometry of the galaxies in  the five $ugriz$ SDSS broad-bands together with some other different bands mimicking Euclid and other surveys are also included. Unfortunately, no information on the original spectrum was kept due to disk space issues. While this fact makes a direct comparison impossible, we can obtain an estimation of their spectral type by fitting the photometry to a library of templates, as detailed in \S3.2.

\subsection{Mock photometry and photometric redshifts with  \texttt{PhotReal}}

While the advantages of these semi-analytic mock catalogues are crearly recognized, well-known issues have been widely reported in the literature related to unrealistic galaxy colours  \citep{cohn07,weinmann11,skelton12,somerville12,hansson12,henriques12}. Many of these issues consequently lead to an overestimation of the mean dispersion of the photometric redshifts \citep{molino14} and flawed stellar mass estimations \citep{mitchell13}. Some of these disagreements for the mock catalogues have been reported in  \cite{merson15} and \cite{ascaso15b}. These issues, together with the lack of photometric errors in the original mock catalogue motivated us to create a new mock catalogue that reproduces the properties of the observed galaxies as well as their photometric redshift precision in all the bands of J-PAS.

In order to do this, we apply \texttt{PhotReal} (\citealt{ascaso15b}, Ben\'itez et al. in prep.). This procedure, already applied in several papers \citep{arnalte-mur14,zandivarez14,ascaso15a,ascaso15b}, ensures the accurate reproduction of the magnitudes, colours and photometric redshifts of the galaxies. In this section, we provide a brief summary of the method. For further details, we refer the reader to  \cite{ascaso15b} and Ben\'itez et al. in prep. 

\texttt{PhotReal} first obtains an estimate of the spectral type of the original catalogue by matching the original rest-frame mock photometry  to a well-calibrated library of galaxy templates. This library includes eight different empirical templates representing a complete representation of the colors of any galaxy populations sampled by J-PAS (Ben\'itez, private communication). Indeed, this library represents the observed properties of the ALHAMBRA and COSMOS surveys with an outlier rate of $\sim$ 1-2\% \citep{rafelski15}.

Once we have a realistic representation of the spectral type distribution of the original mock catalogue, we generate photometry in the different J-PAS bands by using their filter response and with their expected depths \citep{benitez14}. We include an empirically calibrated systematic error of about 7\%. This error remains constant with magnitude and seems to be intrinsic to the galaxy colors in multi-band photometry. Furthermore, photometric and instrumental errors are added to these magnitudes. The photometric errors are estimated as in \cite{benitez09a} from the telescope response and are normalized to the J-PAS depths in each band. Thereafter, we run \texttt{BPZ2.0} on the new photometry to  obtain photometric redshifts and redshift probability distribution functions, $P(z)$.

One might claim that the use of the same library could be introducing an optimistic behavior of the performance of the photometric redshifts. However, we checked that this was not the case in a previous work \citep{ascaso15b}. First of all, we ensured that the photometry generated with \texttt{PhotReal} was very similar to the one observed in the literature, which if any, it should make the derived photometric redshifts more realistic. Secondly, we checked that the photometric redshifts derived from this photometry perfectly matched with the photometric dispersion and bias measurement in real data. For example, we measured the original photometric redshift dispersion of the mock catalogue generated for the Advanced Large, Homogeneous Area Medium Band Redshift Astronomical survey (ALHAMBRA, \citealt{moles08}) data, finding it to be three times higher than the one expected from real data \citep{molino14}. After applying \texttt{PhotReal}, the photometric redshift dispersion exactly matched the data expectations \citep{ascaso15a}. Note that the choice of the library was motivated from an excellent calibration of the template library in representing $>$98\% of the known galaxies. Moreover, possible interpolations between their templates are allowed when obtaining photometric redshifts.

\section{Photometric redshifts}

\subsection{J-PAS photometric redshift predictions}

In this section, we describe the performance of the photometric redshifts estimated for J-PAS obtained from the mock catalog described in Section \S3. Following \cite{brammer08,molino14}, we define the photometric redshift dispersion as:

\begin{equation}
\sigma_{NMAD}=1.48\times \big \langle\bigg(\frac{|\Delta z - \langle \Delta z \rangle|}{1+z_s}\bigg)\big\rangle
\end{equation}
\noindent  where the difference $\Delta z = z_b -z_s$ is defined as the photometric redshift bias (see \citealt{molino14} and references herein) and $z_b$ and $z_s$ refer to the bayesian photometric and the spectroscopic redshifts respectively. 

\begin{figure}
\centering
\includegraphics[clip,angle=0,width=1.0\hsize]{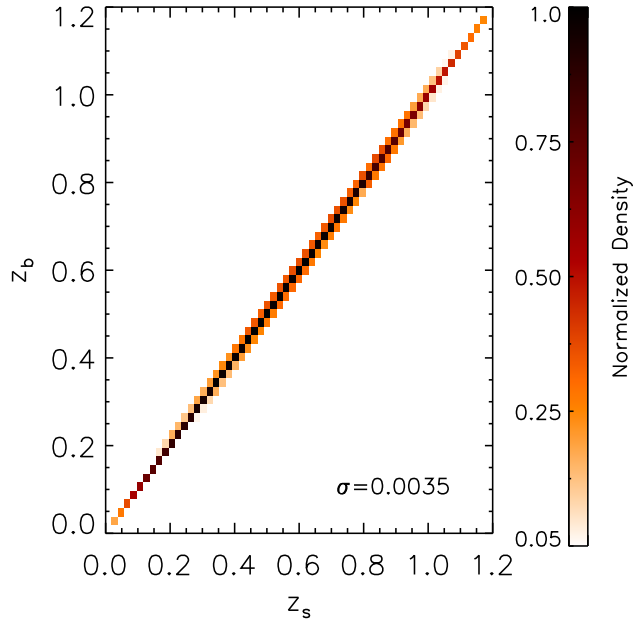} 
\caption{Density plot of the photometric redshift $z_b$ versus spectroscopic redshift $z_s$ for the overall J-PAS sample, colour-coded by normalized density. The photometric redshift dispersion is quoted. The J-PAS photometric redshift will perform with a similar resolution to this obtained for low-resolution spectra up to redshift 1.2.}
\label{fig:zszb}
\end{figure}

\begin{figure}
\centering
\includegraphics[clip,angle=0,width=1.0\hsize]{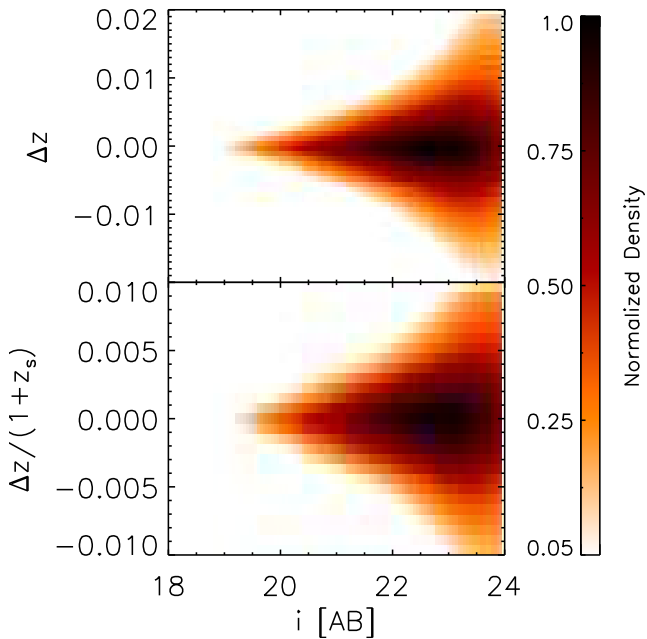} 
\caption{Density plot of the photometric redshift bias (upper panel) and photometric redshift dispersion (bottom panel) as a function of i-band magnitude for the overall J-PAS sample, colour-coded by normalized density. The photometric redshift behavior as a function of the magnitude is several times better than other next-generation surveys (see \S4.2).}
\label{fig:deltazmag}
\end{figure}

\begin{figure}
\centering
\includegraphics[clip,angle=0,width=1.0\hsize]{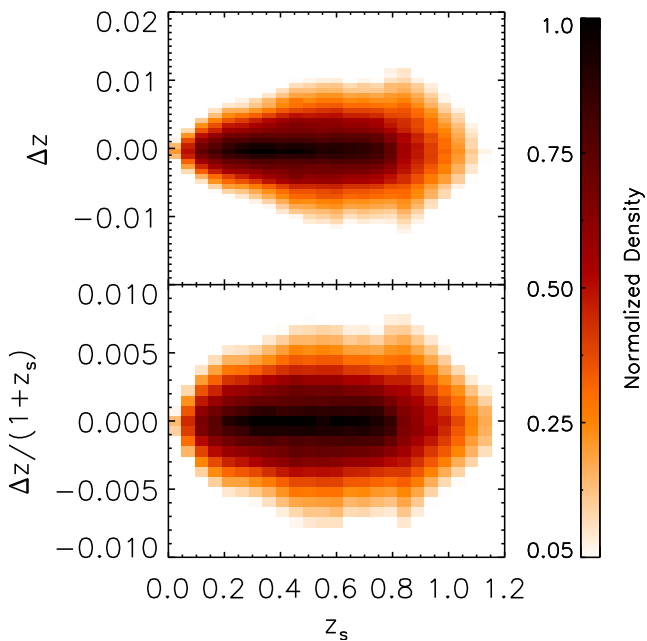} 
\caption{Density plot of the photometric redshift bias (upper panel) and photometric redshift dispersion (bottom panel) as a function of redshift for the overall J-PAS sample, colour-coded by normalized density.}
\label{fig:sigmazz}
\end{figure}

The overall photometric redshift dispersion obtained for the global J-PAS mock sample is $\sigma_{\rm NMAD} = 0.003$, equivalent to $\sim 1000$ km s$^{-1}$.  Fig. \ref{fig:zszb} shows the normalized density plot of the spectroscopic redshift versus photometric redshift for all the sample. It is noticeable the tightness of the relation as expected from the excellent spectrum coverage of J-PAS. In addition, Figs \ref{fig:deltazmag} and \ref{fig:sigmazz} show the density plot of the relation of the photometric redshift bias (upper panel) and dispersion (bottom panel) as a function of magnitude and redshift,  respectively. We complement this information with the first two columns of Tables \ref{tab:photozmag} and \ref{tab:photozz}, showing the photometric redshift precision and photometric redshift bias as a function of magnitude and redshift, respectively.

The photometric redshift bias keeps well below 0.0015 up to $i\sim 23.5$ mag and $z\sim 1.2$, at least. The photometric redshift dispersion remains below 0.003 up to $i\sim 23.0$ mag and up to $z\sim 0.8$. This performance is markedly better  than similar next-generation surveys in the same range of redshift and magnitude  \citep{ascaso15b}, as we discuss in \S4.2,  putting in evidence the very good behavior of a pseudo-spectra-like survey as J-PAS. 

We can argue that the utilized mock catalogues might be too simplistic and therefore unrealistic since they do not include other sources of errors except the photometric errors. However, as seen in previous analysis based on real data such as the ALHAMBRA survey \citep{ascaso15a}, other sources of errors such as those coming, for instance, from the chosen library of templates, has a small impact on the results. Therefore, we are confident that the present results in this work represent a realistic expectation of the capacity of the J-PAS survey.

We have also characterized the rate of outliers expected in the survey. To do this, we have chosen to use two different definitions as in \cite{molino14}. First, we call a galaxy an outlier  if it satisfies the following condition:

\begin{equation}
\frac{|\Delta z|}{1+z_s}>0.15.
\end{equation}
Hence, the rate of outliers, $\eta_1$, is defined as:
\begin{equation}
\eta_1=N\big(\frac{|\Delta z|}{1+z_s}>0.15\big)/N_T
\end{equation}
where $N_T$ refers to the total number of galaxies.

We decided to use this definition since it has been the referred quantity in a large number of papers in the literature for historical reasons (e.g. \citealt{ilbert06,ilbert09,coupon09,hildebrandt10,hildebrandt12,raichoor14}). However, as noted in \cite{hildebrandt12}, this definition becomes kind of arbitrary, particularly for surveys with a large-number of narrow-medium bands. Hence, we also use the definition of an outlier galaxy, as that accomplishing the following condition:

\begin{equation}
\frac{|\Delta z|}{1+z_s}>5 \sigma_{NMAD}
\end{equation}
as already introduced in \cite{molino14}. As before, the rate of outliers, $\eta_2$, is defined as
\begin{equation}
\eta_2=N\big(\frac{|\Delta z|}{1+z_s}>5 \sigma_{NMAD}\big)/N_T
\end{equation}

\begin{figure}
\centering
\includegraphics[clip,angle=0,width=1.0\hsize]{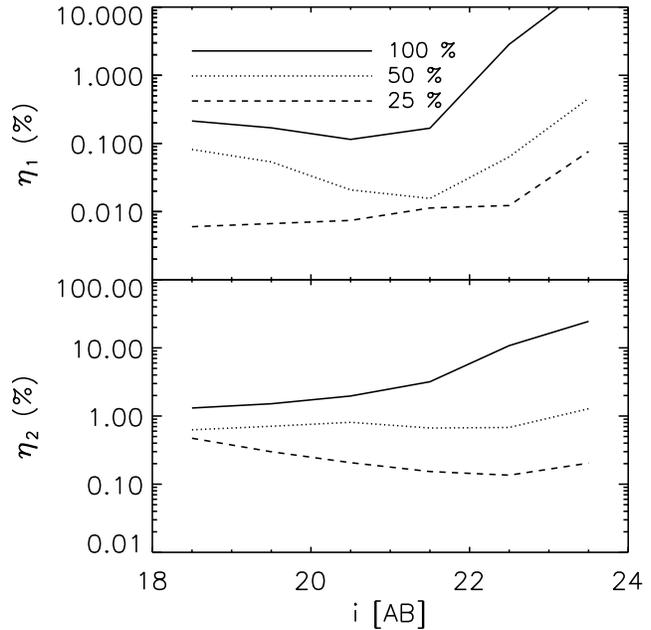} 
\caption{Outliers rate defined as $\eta_1$ (top panel) and $\eta_2$ (bottom panel) as a function of magnitude for three different J-PAS samples: the overall sample (solid line), the 50\% best quality photometric redshift sample (dotted line) and the 25\% best quality photometric redshift sample (dashed line). The outliers rate is very small ($\eta_1<$1\% and $\eta_2<$6.5\%) down to $i\sim22.5$, increasing to higher rates at fainter magnitudes for the overall sample, and for all the 50\% and 25\% best quality J-PAS samples.}
\label{fig:outmag}
\end{figure}

\begin{figure}
\centering
\includegraphics[clip,angle=0,width=1.0\hsize]{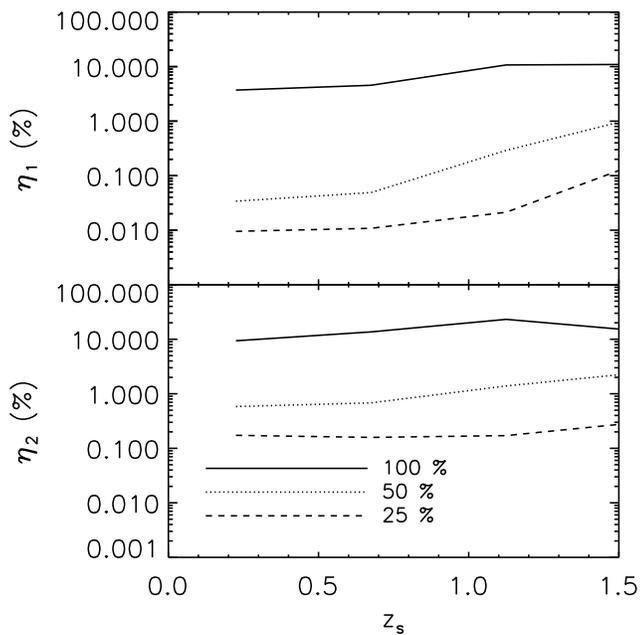} 
\caption{Outliers rate defined as $\eta_1$ (top panel) and $\eta_2$ (bottom panel) as a function of redshift for three different J-PAS samples: the overall sample (solid line), the 50\% best quality photometric redshift sample (dotted line) and the 25\% best quality photometric redshift sample (dashed line). The $\eta_1$ outliers rate keeps below 5\% for the overall sample ($\eta_2<12.5\%$) down to $z\sim0.8$ and almost negligible ($\eta_1$ and $\eta_2<$1\%) for the 50\% and 25\% best quality J-PAS samples.} 
\label{fig:outz}
\end{figure}

The solid lines in Figs. \ref{fig:outmag} and \ref{fig:outz} display the  rate of outliers ($\eta_1$ top panel, $\eta_2$ bottom panel) expected for J-PAS as a function of magnitude and redshift respectively. Also, these values are collected in the third and fourth columns in Tables \ref{tab:photozmag} and \ref{tab:photozz}, respectively. The rate of outliers becomes almost negligible ($\eta_1<$1\% down to i$\sim$ 22.5 and $\eta_2<$3\% down to i$\sim$21.5) and it increases up to $\eta_1<$4\% and up to $\eta_2<$13.7\% down to i$\sim$23. This rate is slightly higher ($\eta_1<$5\% and $\eta_2<$12.5\%) up to z$\sim$0.8, as expected from the inclusion of fainter low-redshift galaxies in this sample. These values are significantly ($\sim$5-20 times) smaller  than the values that other next-generation surveys will achieve \citep{ascaso15b}, in agreement with what it is expected for surveys with a large number of narrow-band filter  \citep{benitez09a}. A more detailed comparison is performed in \S4.2.

Complementarily, \cite{benitez00} introduced an indicator of the quality of the photometric redshift of a survey with a parameter called \emph{odds}. This parameter is defined as the integral of the full redshift probability $P(z)$ centered in the maximum peak of the probability within a given interval:

\begin{equation}
odds=\int_{z_b-2\sigma_{NMAD}}^{z_b+2\sigma_{NMAD}}p(z) dz,
\end{equation}
\noindent and it becomes a very useful quantity to select the best quality photo-z samples. 

\begin{figure}
\centering
\includegraphics[clip,angle=0,width=1.0\hsize]{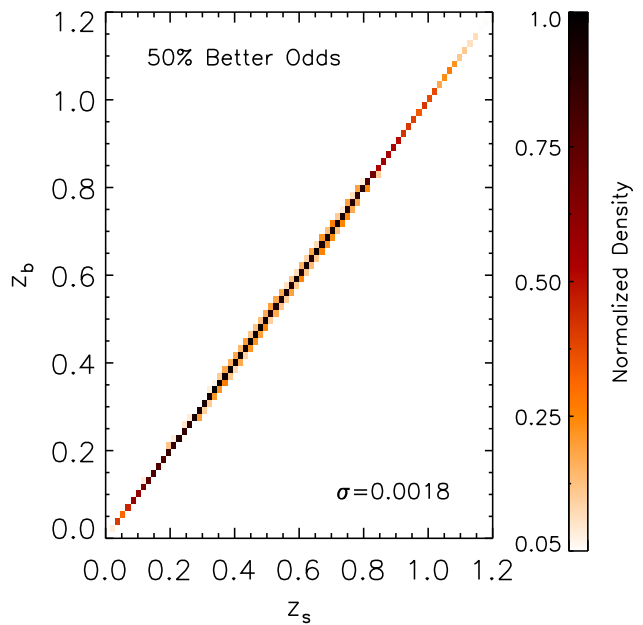} 
\caption{Density plot of the photometric redshift $z_b$ versus spectroscopic redshift $z_s$ for the J-PAS sample resulting after selecting the 50\% best quality photometric redshifts in the sample, selected through the \emph{odds} parameter (see the text for an explanation), colour-coded by normalized density. The photometric redshift dispersion is quoted. This dispersion is comparable to that obtained for a low-resolution spectroscopic survey.}
\label{fig:zszbOdd}
\end{figure}

In  Fig. \ref{fig:zszbOdd}, we show the recovery of photometric redshifts for the J-PAS survey for the 50\% best quality \emph{odds} sample, which have a  mean dispersion of 0.0018. We also display the  rate of outliers for the 50\% and 25\% best \emph{odds} samples as a function of magnitude and redshift in Figs \ref{fig:outmag} and \ref{fig:outz} respectively. Likewise, in Tables \ref{tab:photozmag} and \ref{tab:photozz}, we list the photometric redshift precision, the photometric redshift bias and the rate of outliers, $\eta_1$ and $\eta_2$, as a function of magnitude and redshift, respectively, for different \emph{odds} cuts. 

While the global photometric redshift dispersion, $\sigma_{NMAD}$ remains below 0.003 down to $i\sim23.0$, performing a selection of the 50\% highest  \emph{odds}  allows the mean photometric redshift dispersion to decrease down to 0.0015 or to even smaller values down to $i\sim24$. In the same way, the overall outliers rate, $\eta_1$ ($\eta_2$), remains below 1\% (6.3\%) for magnitude $i<22.5$ mag but it increases to $>$4\% ($>$14\%) at deeper magnitudes than this. An \emph{odds} cut leaving 50\% of the sample, automatically reduces the outliers rate to $\eta_1<$1\% and $\eta_2<$2\% down to $i\sim23.5$ and to $\eta_1<$0.25\% and $\eta_2<$0.5\% down to $i\sim24$ if we consider the 25\% best quality \emph{odds} galaxies in the sample.

\begin{table*}
      \caption{J-PAS photometric redshift bias, photometric redshift dispersion and rate of outliers as a function of magnitude for different \emph{odds} cut.}
      \[
         \begin{array}{|c|cccc|cccc|cccc|}
		\hline
		{mag} &  {\rm } & {\rm 100\%} & {\rm best \, odds} &    & & {\rm 50\%} & {\rm best \, odds}&  &   &{\rm 25\%} & {\rm best \, odds} &     \\
		{} &  {\rm \Delta z} &  {\rm \sigma_{NMAD}} & {\rm \eta_1 (\%)} & {\rm \eta_2 (\%)} &   {\rm \Delta z} &   {\rm \sigma_{NMAD}} &{\rm \eta_1 (\%)}   &{\rm \eta_2 (\%)} &  {\rm \Delta z} &{\rm \sigma_{NMAD}} &   {\rm \eta_1 (\%)}  &   {\rm \eta_2 (\%)} \\\hline
     18.0 - 18.5 &    0.0001  &    0.0015 &     0.22  &  1.22 &   0.0001  &    0.0013	&  0.08  & 0.55   & 0.0000  &    0.0009 & 0.01 &  0.52 \\
     18.5 - 19.0 &    0.0000 &     0.0014  &    0.21   &	 1.37  & 0.0000  &    0.0012   & 0.08  & 0.65   &  0.0000 &     0.0008 &	 0.01  &  0.47 \\
     19.0 - 19.5 &    0.0000 &     0.0014  &    0.19   &	 1.44 & 0.0620  &    0.0012	&  0.06   & 0.67 &  0.0000  &    0.0008 &	 0.01 & 0.36 \\  
     19.5 - 20.0 &    0.0000  &    0.0014 &     0.16   &	 1.57 &  0.0000  &    0.0011	&  0.05   & 0.72 &  0.0000 &     0.0008 &	 0.01 & 0.26 \\   
     20.0 - 20.5 &    0.0000  &    0.0013 &     0.13  & 	 1.77 & 0.0000  &    0.0011 &   0.03   & 0.80 & 0.0000  &    0.0008 &	 0.01 & 0.24 \\   
     20.5 - 21.0 &    0.0000  &    0.0013  &    0.11  & 	 2.08 & 0.0000  &    0.0011 &   0.02   & 0.82 &  0.0000  &    0.0008 &	 0.01 & 0.19\\   
     21.0 - 21.5 &    0.0000  &    0.0014  &    0.11  & 	 2.61 & 0.0000  &    0.0011 &   0.02   & 0.73 & -0.0001  &    0.0008 &	 0.01 & 0.17 \\   
     21.5 - 22.0 &    0.0000  &    0.0015 &     0.21  & 	 3.54 & 0.0000  &    0.0011 &   0.02   & 0.63 & -0.0001 &     0.0009 &	 0.01 & 0.14\\   
     22.0 - 22.5 &    0.0000 &     0.0018 &     0.93  & 	 6.28 & 0.0000  &    0.0013 &   0.04   & 0.63 & -0.0001 &     0.0010 &	 0.01 & 0.14\\   
     22.5 - 23.0 &    0.0001 &     0.0024 &     4.15  & 	 13.67 & 0.0000  &    0.0014	&  0.09  & 0.73 & -0.0001 &     0.0010 &	 0.01 & 0.13\\   
     23.0 - 23.5 &    0.0011 &     0.0050 &     12.65  &  26.60 &  0.0000  &    0.0014 &   0.20   & 0.93 & -0.0001 &     0.0010 &	 0.01 & 0.12\\   
     23.5 - 24.0 &    0.0103 &     0.0414 &     27.09  & 14.89 & 0.0000  &    0.0015	&  1.05   & 2.06 & -0.0001 &     0.0011 &	 0.26 & 0.43\\ \hline 
	\end{array}
      \]
\label{tab:photozmag}
   \end{table*}

Similarly, the overall photometric redshift resolution remains below $\sigma_{NMAD}<0.003$ up to redshift 0.8, increasing at higher redshifts. However, by selecting the 50\% best quality \emph{odds} sample, these values remain $<$0.0015 up to redshift 1.0. More strikingly, while the $\eta_1$ outlier rates range between 3\% and 11\% and $\eta_2$ between 8\% and 23\% for the whole redshift range, selecting the best 50\% of the sample can decrease these assessments to less than 1\% and 2\% for $\eta_1$ and $\eta_2$ respectively. 
   
\begin{table*}
      \caption{J-PAS photometric redshift bias, photometric redshift dispersion and rate of outliers as a function of redshift for different \emph{odds} cut.}
      \[
         \begin{array}{|c|cccc|cccc|cccc|}
		\hline
		{z_s} &  {\rm } & {\rm 100\%} &{\rm best \, odds}   &  & &  {\rm 50\%} & {\rm best \, odds}&   &  &{\rm 25\%} & {\rm best \, odds}&      \\
		{} &  {\rm \Delta z} &  {\rm \sigma_{NMAD}} & {\rm \eta_1 (\%)} & {\rm \eta_2 (\%)} & {\rm \Delta z} &   {\rm \sigma_{NMAD}} & {\rm \eta_1 (\%)}  & {\rm \eta_2 (\%)} &{\rm \Delta z} &    {\rm \sigma_{NMAD}} & {\rm \eta_1 (\%)} & {\rm \eta_2 (\%)}   \\\hline
      0.0 - 0.2  &    0.0001  &    0.0016   &   4.91  & 8.69   &	 0.0000    &  0.0012	& 0.06  & 0.56  &     0.0000  &    0.0010 &	    0.02   & 0.17  \\
      0.2 - 0.4  &    0.0000  &    0.0017   &   3.41    &	9.26 & 0.0000    &  0.0011	& 0.03   & 0.59 &    -0.0001  &    0.0009 &	0.01  & 0.17  \\     
      0.4 - 0.6  &    -0.0001 &     0.0021  &    3.31  &  	11.50 & 0.0000    &  0.0013	&  0.01  & 0.60 &     -0.0001 &     0.0009 &	0.01   & 0.18  \\
      0.6 - 0.8  &     0.0000 &     0.0021  &    3.70   & 	12.40 & 0.0000    &  0.0012	& 0.04  &  0.65 &	0.0000  &    0.0009 &	0.01   & 0.14  \\
      0.8 - 1.0  &    0.0016  &    0.0037   &  10.38   & 	21.24 & -0.0002    &  0.0014	&  0.25  &   1.11 &   -0.0002 &     0.0010 &	0.04    & 0.21 \\
      1.0 - 1.2  &    0.0024  &    0.0044  &    10.16   & 23.02 & 0.0000    &  0.0016	&  0.24  &  1.41 &    -0.0001 &     0.0011 &	0.01   & 0.14  \\
      1.2 - 1.4  &    0.0013  &    0.0035  &    7.02   & 23.01 &	 0.0000    &  0.0015	&  0.25  &   1.15 &   -0.0001 &     0.0011 &	0.04   & 0.15  \\ \hline
	\end{array}
      \]
\label{tab:photozz}
   \end{table*}

\subsection{Comparison with Euclid and the LSST surveys}

We compare our results with those shown in \cite{ascaso15b} for the LSST and Euclid surveys.  The authors considered two Euclid surveys consisting of the three infrared $YJH$ Euclid bands and two different optical counterparts. The so-called Euclid-Pessimistic includes the five $grizy$ DES optical bands as the optical counterpart and the Euclid-Optimistic includes the previously mentioned five optical DES bands and the 6 $ugrizy$ LSST deep optical bands. For further details on these surveys, we refer the reader to \cite{ascaso15b}.

\begin{figure*}
\centering
\includegraphics[clip,angle=0,width=0.75\hsize]{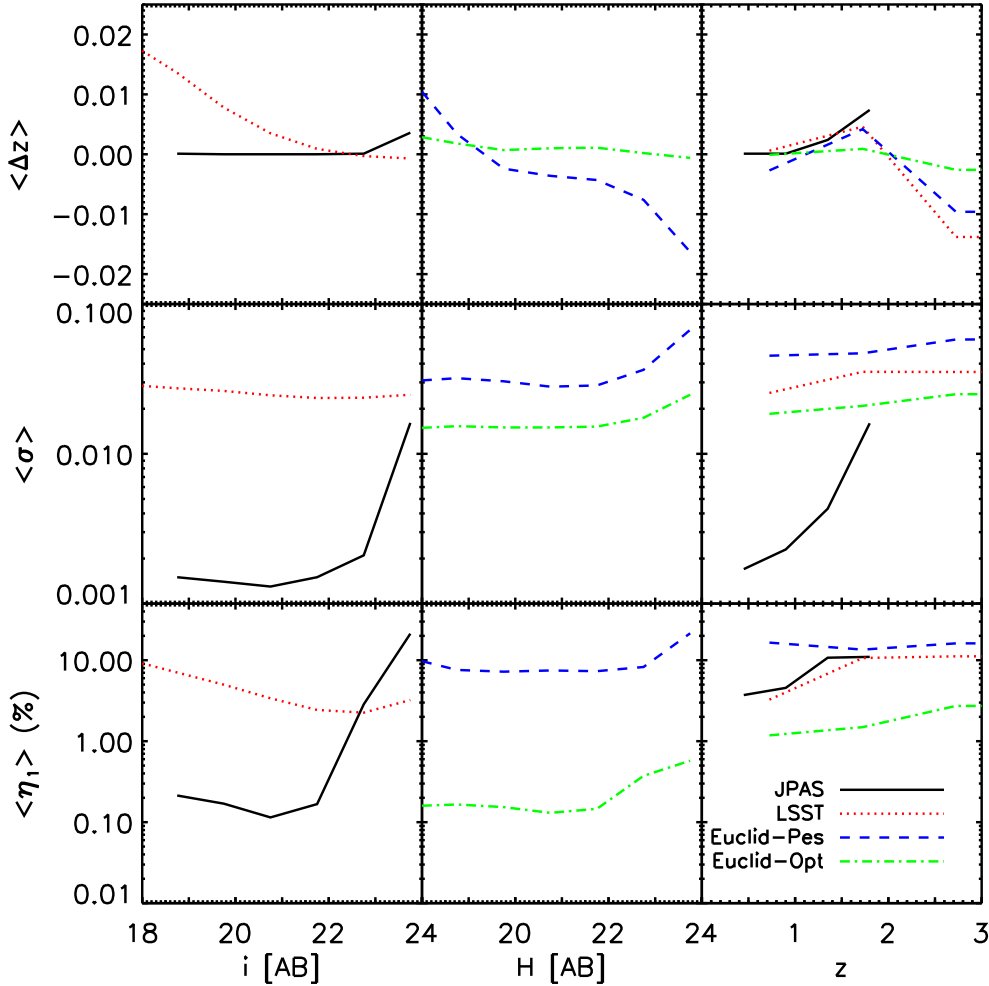} 
\caption{Mean photometric redshift properties for four different next-generation surveys: the J-PAS survey (black solid line), the LSST survey (red dotted line), the Euclid-Pessimistic survey (blue dashed line) and the Euclid-Optimistic survey (green dotted-dashed line). 
From top to bottom, the mean photometric redshift bias, the mean photometric redshift dispersion and the mean outliers rates are plotted as a function of the i-band magnitude (for the J-PAS and LSST surveys), H-band magnitude (for the Euclid surveys) and redshift, from left to right. For the dependence of the different properties as a function of redshift, the mock catalogues are restricted down to $i=$23.5 (J-PAS), $i=$27.0 (LSST) and $H=$24 (Euclid).}
\label{fig:comparisonPhotoz}
\end{figure*}

In order to be consistent in our comparison and to use the same definition of outlier, we only use $\eta_1$ in this subsection and refer to it as $\eta$. In Fig. \ref{fig:comparisonPhotoz}, we show the mean photometric redshift bias, the mean photometric redshift dispersion and the mean outlier rate (top, middle and bottom panel, respectively) as a function of the i-band magnitude (for the J-PAS and LSST surveys), H-band magnitude (for the Euclid surveys) and redshift (left, centre and right panel, respectively).

We find that the photometric redshift bias as a function of the i-band magnitude obtained for J-PAS is $>$100 times smaller than those found for the LSST.  The values found for the Euclid-Optimistic as a function of the H-band magnitude are $>$5 times smaller than those found for the Euclid-Pessimistic. Comparing the photometric redshift bias as a function of redshift, the values found for J-PAS are a factor 1-20 smaller than those found for the  LSST and Euclid-Pessimistic and in the same range of magnitude as those found for the Euclid-Optimistic survey.

The mean photometric redshift dispersion for the J-PAS survey is more than a factor of 20 smaller than the LSST and Euclid-Pessimistic survey, and more than a factor of 10 smaller for the Euclid-Optimistic survey for similar magnitudes ranges. These differences remain similar, although slightly smaller, when we consider it as a function of redshift, up to redshift $<$1. At higher redshifts, the photometric redshift dispersion increases almost exponentially for the J-PAS survey.

Finally, the rate of outliers for the J-PAS survey as a function of magnitude is more than 15 times smaller than for the LSST and Euclid-Pessimistic and similar to the Euclid-Optimistic survey down to $i\sim22.5$. However, the J-PAS outliers rate becomes comparable to the LSST values ($\sim$ 3-5\%) as a function of redshift and higher than the ones expected for the Euclid-Optimistic survey. 

The results discussed in this section illustrate that the different observational strategies of the next-generations surveys provide different photometric redshift performances. For instance, deep IR surveys such as Euclid will allow to reach high redshift regimes and very deep optical surveys, such as the LSST, will sample a wide range of the luminosity function, reaching very deep magnitudes. However, as shown in \cite{benitez09b}, although somewhat counterintuitive, medium and narrow band filter systems produce much more robust photometric redshifts, and there much larger photometric redshift depth, than broad-band systems  reaching higher S/N. Indeed, only the surveys that use the combination of multiple bands (e.g J-PAS or Euclid-Optimistic) will reach very small bias and outliers rates. In addition, if these bands are narrow and cover the whole optical spectrum as J-PAS will do, the photometric redshift dispersion will be reduced to very low-levels down to the depth limit of the survey.

\section{J-PAS galaxy cluster survey}

\subsection{The Bayesian Cluster Finder}

The Bayesian Cluster Finder, (BCF, \citealt{ascaso12,ascaso14a,ascaso15a}) is an optical/IR galaxy cluster detector, which was developed with the purpose of detecting galaxy clusters independently of the presence or absence of a red sequence or a central dominant brightest cluster galaxy (BCG) but using this information if it is present. In other words, the algorithm uses the presence of a red sequence or a dominant BCG if present, but it can still detect galaxy clusters or groups if this information is absent. In this section, we summarise the main details of the method and we refer the reader to the original work for further details. 

The BCF first calculates the probability at a given redshift that there is a cluster with a determined density and luminosity profile centered on each galaxy, including different priors related to the colour-magnitude relation of the cluster or the BCG magnitude-redshift relation. Then, we perform a search in a predefined number of redshift slices, where the minimum threshold comes from the minimum redshift we can resolve (usually determined from the geometry of the survey for small area surveys or 0, otherwise) and the maximum redshift is obtained from the wavelength coverage and the depth of the survey. The bin width is fixed according to the expected photometric resolution of the survey. For instance, we fix the photometric redshift resolution to 0.01 for the J-PAS survey. 

Effects of stars and masking of edges of the frames have been incorporated. Then, clusters are selected as the density peaks of those probability maps and the center is located at the peak of the probability. Finally, if we find two or more detections with separations less than 0.5 Mpc in projected space and up to two bins in redshift space, we merge them into a single one.

The BCF has been applied to a number of optical surveys in the literature: a wide (141 $deg^2$) survey, the CFHTLS-Archive Research Survey (CARS, \citealt{erben09}; \citealt{ascaso12}); a very deep (r$\sim$27.5 mag depth) survey, the Deep Lens Survey (DLS, \citealt{wittman02}; \citealt{ascaso14a}); and a 20 medium-band survey, the ALHAMBRA survey (\citealt{moles08,ascaso15a}). We remark different  performances as a function of the different properties of the data. Particularly, we see that multiple narrow-band surveys such as the ALHAMBRA survey are better at resolving the galactic population and therefore, at increasing the purity rates and setting a lower mass limit threshold for detecting clusters and groups.

\subsection{Selection function}

In this work, we apply the BCF to our J-PAS mock catalogue.  In order to assess the performance of the BCF on the J-PAS mock data, we  match the original mock sample to the recovered sample following the same Friends-of-Friends (FoF, \citealt{huchra82}) algorithm described in \cite{ascaso12}. This procedure searches for each detection found in the recovered sample candidates or 'friends' in the original mock catalogue whose centers are placed within a comoving distance of 3 Mpc including the photometric redshift errors. Then, a search of FoF is done until no more candidates are found. Afterwards, the candidate with the closest photometric redshift to the original detection is selected. Finally, if this detection is found within a distance of 1 Mpc, we consider this to be a match.

We compute our observable richness, the total stellar mass, $M^*_{\rm CL}$, defined as the sum of stellar mass of all the galaxies belonging to the clusters brighter than the magnitude limit within a certain radius \citep{ascaso15a}. This observable has been chosen  instead of the $\Lambda_{CL}$ used in other work \citep{postman02,ascaso12,ascaso14a} or $N_{200}$ used usually in red sequence based methods as it has been proved one the optical measurables with smaller intrinsic scatter with the halo mass (e.g \citealt{gonzalez07,andreon10,andreon12,ascaso15a}). In future work, we will explore the possibility of using the photometric redshift functions, $P(z)$, to obtain proxies for the kinematical mass of the cluster.

\begin{figure}
\centering
\includegraphics[clip,angle=0,width=1.0\hsize]{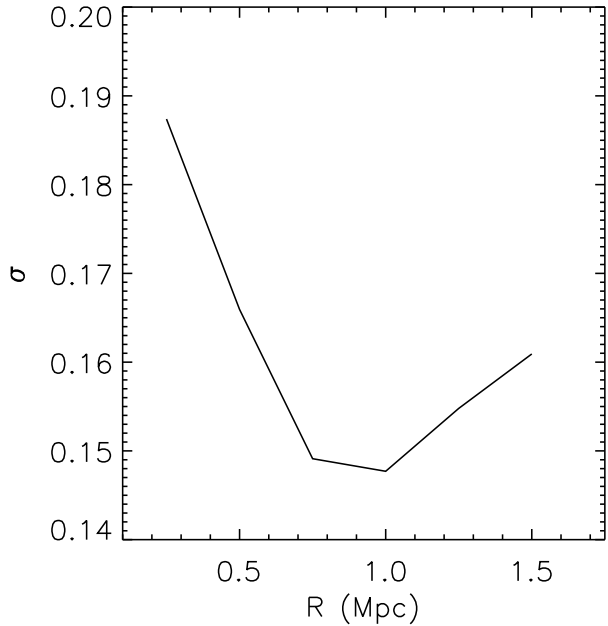} 
\caption{Main scatter $\sigma_{M^*_{\rm CL} | M_{\rm h}}$, between the total stellar mass $M^*_{\rm CL}$ and the mass halo, $M_{\rm h}$, as a function of different considered radius to compute $M^*_{\rm CL}$. The minimum scatter is obtained for a $R=1$ Mpc for this data.}
\label{fig:radiusmstellar}
\end{figure}

In order to choose the optimal aperture to compute the total stellar mass for the data presented in this work, we chose six different apertures ranging from 0.5 to 1.5 Mpc in steps of 0.25 Mpc. We fit the $M^*_{\rm CL}$- $M_{\rm h}$ relation following the procedure that it is described in section \S5.3 and we computed the main scatter,  $\sigma_{M^*_{\rm CL} | M_{\rm h}}$,  between these two variables. In Fig. \ref{fig:radiusmstellar}, we show the main measured scatter obtained for the different radius. The radius for which the minimum scatter is achieved is $R=1 Mpc$. Therefore, we adopt this radius for computing the $M^*_{\rm CL}$.

\begin{figure}
\centering
\includegraphics[clip,angle=0,width=1.0\hsize]{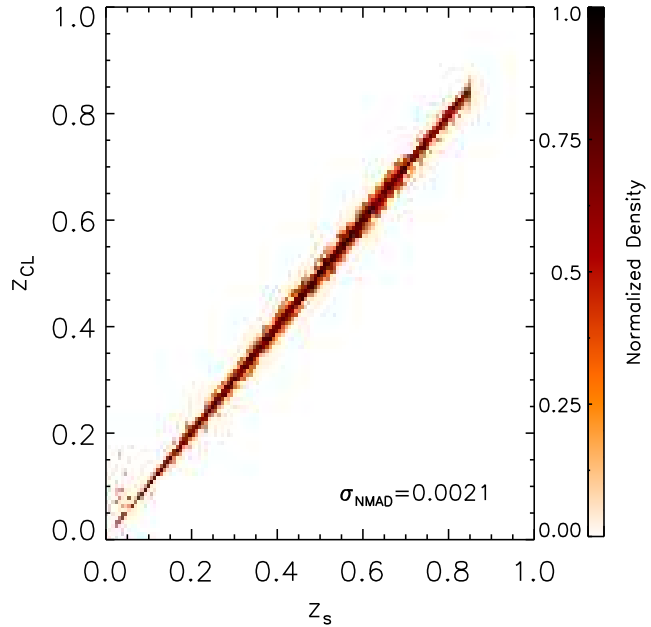} 
\caption{Density map of the cluster photometric redshift estimation and the cluster spectroscopic redshift derived from the mock catalogue for the cluster sample selected by J-PAS. The photometric redshift dispersion is quoted.}
\label{fig:photozBCF}
\end{figure}

We have also investigated which is photometric redshift performance of the BCF cluster finder for the J-PAS clusters. In Fig. \ref{fig:photozBCF}, we show the density map of the cluster photometric redshift estimation versus the cluster spectroscopic redshift for the sample of clusters selected with J-PAS. As we see the sequence is really tight, being the mean dispersion $\sigma_{\rm NMAD}=0.0021$, comparable to the individual accuracy of photo-z's of the bright red galaxies. This result is comparable to the photometric redshift accuracy  found in \cite{rozo14} between optical and X-ray center  up to $z<0.5$ with a negligible percentage of outliers ($<$5\% with a normalized redshift difference $>5 \sigma_{NMAD}$) and it is also significantly better than obtained by other works (e.g.  \citealt{andreon12b}). This outcome illustrates the nice performance of the BCF on recovering the main redshift of the main structures detected with J-PAS.

\begin{figure}
\centering
\includegraphics[clip,angle=0,width=1.0\hsize]{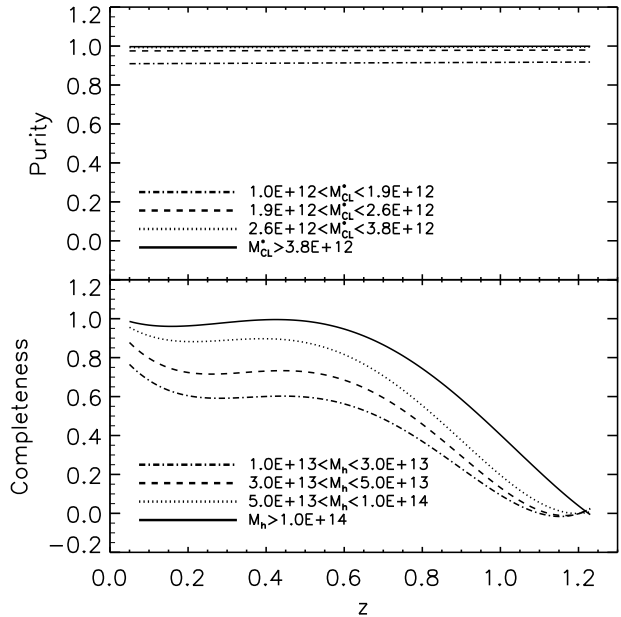} 
\caption{Purity (top plot) and completeness (bottom plot) rates as a function of redshift for different dark matter halo mass ($M_{\rm h}$) and their equivalent total stellar mass ($M^*_{\rm CL}$) bins. The plotted lines have been smoothed by linear (top panel) and fourth order polynomial (bottom panel) interpolation. While purity remains almost constant  as a function of redshift, being lower for lower masses, we find a decreasing trend in the completeness rate with both redshift and mass. According to these rates, we expect to find reliably ($>$80\% completeness and purity) galaxy clusters and groups with total masses down to $M_{\rm h} \sim5 \times 10^{13}M_{\odot}$ up to z$\sim$0.7 and $>$70\% completeness rates down to $M_{\rm h} \sim3 \times 10^{13}M_{\odot}$ up to the same redshift.}
\label{fig:drate}
\end{figure}

Furthermore, we have computed the completeness and purity of the results as a function of redshift and richness.  We define completeness as the rate of clusters detected with the BCF out of the total simulated sample, and the purity as the rate of clusters simulated that were detected  with the BCF out of the total detected sample. In Fig. \ref{fig:drate}, we show the completeness and purity rates as a function of redshift for different values of total stellar mass, $M^*_{\rm CL}$. Their equivalent halo mass $M_{\rm h}$ bins have been obtained through the calibration described in \S5.3 and are also quoted in the bottom panel.

The results  show that we will be able to detect galaxy clusters with completeness and purity rates $> 80$\% for clusters and groups down to $M_{\rm h}\sim5 \times 10^{13} M_{\odot}$ up to redshift 0.7. At higher redshifts, the completeness rates decay, so that we can detect clusters down to  $M_{\rm h}\sim1 \times 10^{14} M_{\odot}$ up to redshift 0.85 with completeness rates higher than $> 80$\%. If instead, we relaxed the completeness rates to be $> 70$\%, the BCF would be able to detect clusters and groups down to $M_{\rm h}\sim3 \times 10^{13} M_{\odot}$ up to redshift 0.7; down to $M_{\rm h}\sim5 \times 10^{13} M_{\odot}$ up to redshift 0.8 and down to $M_{\rm h}\sim1 \times 10^{14} M_{\odot}$ up to redshift 0.9. In this work, we choose to work with the  most conservative threshold of both completeness and purity rates $>$80\%. We then define the clusters selection function as the predicted minimum halo mass threshold for which we can detect galaxy clusters and groups with a completeness and purity rates $> 80\%$ as a function of redshift.  For J-PAS, the expected selection function is constant with $M_{\rm h} = 5\times10^{13}M_{\odot}$ up to z = 0.7 and progressively increases at higher redshifts (see Fig. \ref{fig:drate}).  

Modified FoF algorithms have also been explored to detect galaxy clusters in J-PAS-like narrow-band surveys \citep{zandivarez14}, obtaining lower completeness and purity rates. This is expected since the performance of the FoF algorithms decays for non-spectroscopic surveys due to their sensibility to the linking length. The results with the BCF suffer less contamination and therefore achieve higher completeness and purity rates for the same mass threshold. This result shows the benefit of using cluster finders that use other information either than the spatial to detect galaxy clusters.

It is important to note that the extremely good quality of the photometric redshifts in the J-PAS survey make these results comparable to what we would expect for a low-resolution spectroscopic survey. Indeed, cluster and group searches in spectroscopic surveys such as the VIMOS VLT Deep Survey (VVDS, \citealt{lefevre05}), where a search using Voronoi-Delaunay Tessellation techniques was performed \citep{cucciati10} or the DEEP2 Survey where a group catalogue was obtained based on the Voronoi-Delaunay technique \citep{gerke05}  and another one based on the FoF algorithm  \citep{liu08} found that they could obtain high purity and completeness rates for a similar threshold and with cluster velocity dispersion of $\sim 300-350 \rm km \, s^{-1}$, equivalent to $M_{\rm h} \sim 3-7 \times 10^{13}$ \citep{munari13}.

\begin{figure}
\centering
\includegraphics[clip,angle=0,width=1.0\hsize]{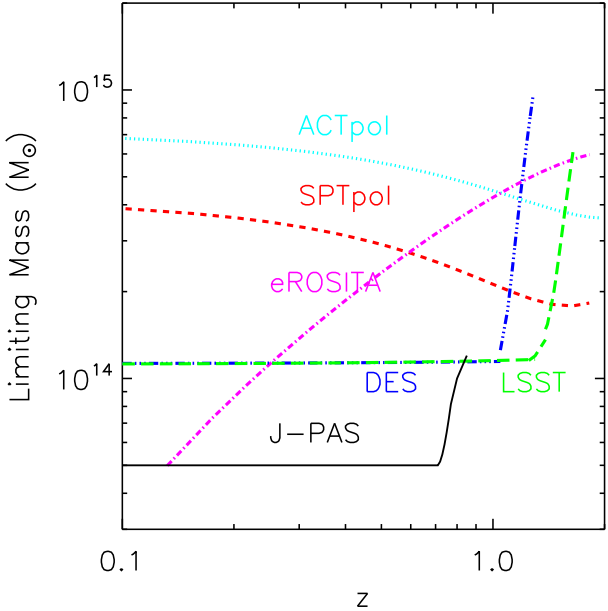} 
\caption{Selection function (minimum mass threshold as a function of redshift) for different next-generation surveys: J-PAS (black solid line), DES (blue three dot-dashed line), LSST (green long dashed line), SPTpol (red short dashed line) and ACTpol (dotted cyan line). The J-PAS becomes the photometric survey reaching the wider range of mass up to $z\sim0.7$.}
\label{fig:selectionFnextG}
\end{figure}

In order to illustrate the enormous benefits of using narrow-band surveys in terms of producing cluster catalogs, we show in Fig. \ref{fig:selectionFnextG} a comparison between the cluster selection function of different next-generation surveys using different observational techniques: X-ray (eROSITA, \citealt{merloni12}), Sunyaev-Zel'dovich (ACTpol, \citealt{niemack10} and SPTpol, \citealt{austermann12}) and optical surveys (DES, \citealt{des05}; LSST, \citealt{lsst09} and J-PAS). All these functions, with the exception of the J-PAS, have been extracted with \texttt{Dexter}\footnote{http://dexter.sourceforge.net/} from \cite{weinberg13}. From this figure, we can clearly see that the selection functions of the optical surveys, while having very similar shapes, also show a large offset with respect to the J-PAS. The 'knee' of the curve is starting at z$\sim$ 0.7 for the J-PAS survey, whereas for DES it happens at $z\sim 1$ and for the LSST at $z>1$. This behavior is related with the depth of the different surveys ($i\sim22.5$ for J-PAS, $i\sim 24.0$ for DES and $i\sim 26.8$ for the LSST). The X-ray eROSITA selection function shows an increasing mass threshold as a function of redshift, obtaining similar mass-groups at low redshift as the J-PAS. On the contrary, the cluster selection functions obtained from the SZ cluster samples show a decreasing lower mass threshold as a function of redshift.

\begin{figure}
\centering
\includegraphics[clip,angle=0,width=1.0\hsize]{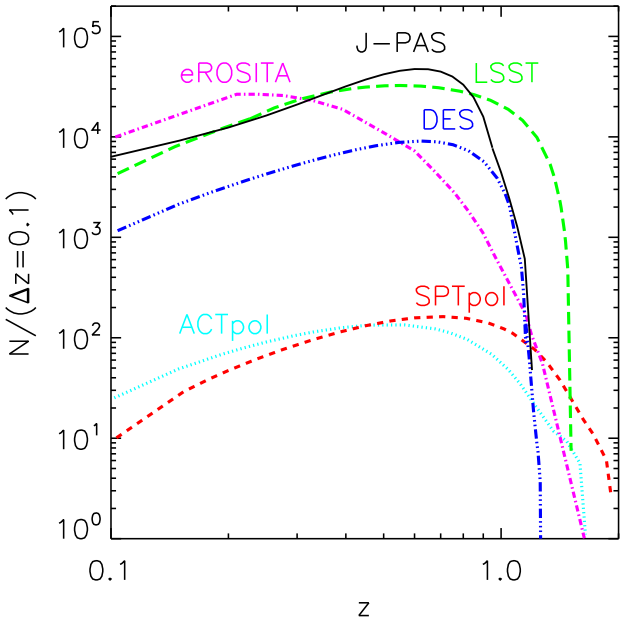} 
\caption{Total number of groups/clusters per redshift bin as a function of redshift for different next-generation surveys: J-PAS (black solid line), DES (blue three dot-dashed line), LSST (green long dashed line), SPTpol (red short dashed line) and ACTpol (dotted cyan line). The J-PAS will detect similar number of clusters and groups as the LSST and eROSITA up to z$\sim$0.7, at least.}
\label{fig:nclustersnextG}
\end{figure}

The impact of the previously shown J-PAS selection function can be seen in Fig. \ref{fig:nclustersnextG}, where we plot the \emph{total number of clusters} as a function of redshift that each survey will observe. As in Fig. \ref{fig:selectionFnextG}, the X-ray and SZ curves have been taken from  \cite{weinberg13}. According to this figure, the number of bound structures detected by J-PAS will be comparable to those found by LSST and eROSITA at least, up to redshift $\sim$0.7 and ten times superior to those found by DES. While the latter surveys will sample with more number statistics the high-end of the mass function, J-PAS will sample the mass function within a wider range of masses. 

Complementarily, the DES and LSST surveys will image a substantial part of the southern sky, whereas J-PAS will provide an optical counterpart of many of the clusters/groups in common with eROSITA in the northern hemisphere up to z=0.7 and some of the most massive higher redshift clusters between $0.7\le z \le 0.85$. This will create an important synergy between the different next-generation surveys that will become very useful for a number of purposes, such as, for instance cosmological purposes.

\subsection{Observable-dark matter halo mass relation}

The optical observable-dark matter halo mass is a crucial relation for cosmological purposes since it allows us to translate an optical measurement into a physical cluster mass (e.g. \citealt{lima05} and references herein). While several efforts have been invested in probing that optical cluster mass tracers can achieve accuracies similar to SZ or X-ray tracers, so far it only has been probed up to moderate redshift \citep{andreon10} or massive clusters \citep{andreon12,saro15}.

In this section, we empirically calibrate the total stellar mass observable-theoretical dark matter halo mass relation, $M^*_{\rm CL} | M_{\rm h}$, from the J-PAS simulations. The fact that the observable used in this work, the total stellar mass, $M^*_{\rm CL}$, is defined down to the flux limit where the survey is complete prevents us from introducing any bias up to the redshift limit where the survey is complete ($z\sim0.7$ for J-PAS).

Inspired by different works (e.g. \citealt{lin06,andreon10,andreon14,saro15}), we have model the $M^*_{\rm CL} | M_{\rm h}$ relation with a log-log relation as follows:

\begin{equation}
<\log M^*_{\rm CL} | M_{\rm h},z>=p_0+p_1\log \big( \frac{M_{\rm h}}{M_{\rm pivot} (M_{\odot})}\big)+ p_2 \log (1+z)
\label{eq:fitMM}
\end{equation}
where $\log$ refers to the decimal logarithmic, $z$ is the redshift of the cluster and $p_i$ are free parameters.  We choose $M_{\rm pivot}=5\times 10^{13}M_{\odot}$ as a reasonable value that represents the expected cluster population. We have fit our data restricted to $z\le 0.75$ and $M\ge 5\times 10^{13}M_{\odot}$  to this model by using an iterative non-linear least-squares minimization method based on the Levenberg-Marquardt algorithm \citep{press92}.  We performed a Monte Carlo simulation sampling 8000 different initial values to compute the fit. The best fitting parameters for the model, together with their 68\% confidence level are listed in Table \ref{tab:fitMM}. Note that the results of this fit have been used to obtain the completeness and purity curves for different observable $M^*_{\rm CL}$ in \S5.2.

\begin{table}
      \caption{Best fitting parameters of the function (\ref{eq:fitMM}) together with their 68\% confidence level.}
      \[
         \begin{array}{cc}
		\hline
		{\rm Parameter} &  {\rm Best \, fit}     \\\hline
p_0 & 12.414 \pm 0.002 \\
p_1 & 0.566 \pm 0.054 \\
p_2 & -0.001 \pm 0.002 \\\hline
\sigma_{M^*_{\rm CL} | M_{\rm h},z} & 0.142 \quad {\rm dex} \\ \hline 
	\end{array}
      \]
\label{tab:fitMM}
   \end{table}

\begin{figure}
\centering
\includegraphics[clip,angle=0,width=1.0\hsize]{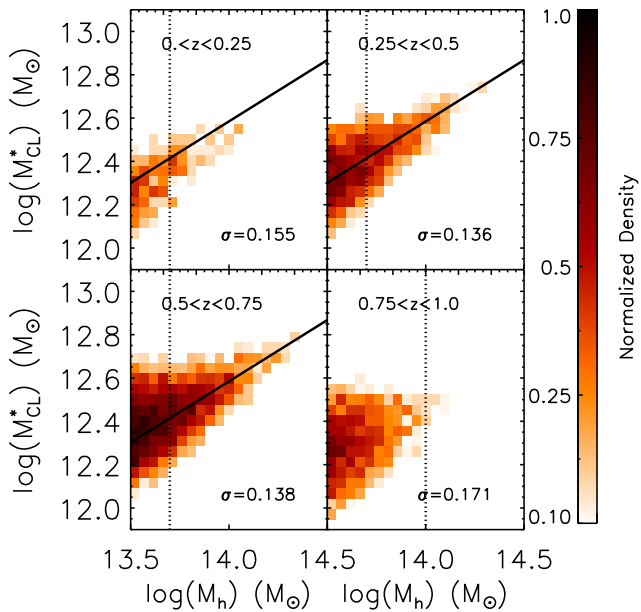} 
\caption{Density plots of the logarithm of the total stellar mass in the cluster as a function of the logarithm of the dark matter mass halo for the matched clusters in the J-PAS mock catalogue for different redshift bins. The solid line indicates the linear fit obtained down to $M_{\rm h}=5\times10^{13}M_{\odot}$ in the first three redshift bins. Each panel shows the scatter measured for each different redshift bin.}
\label{fig:mhalolamda}
\end{figure}

The $M^*_{\rm CL} | M_{\rm h}$ relation  appears not to evolve with redshift, in agreement with other works \citep{lin06,andreon14,saro15}. In Fig. \ref{fig:mhalolamda}, we show the density plot of the relation between the total stellar mass parameter and the dark matter halo mass for different redshift bins. The solid line shows the fit for a particular redshift bin up to $z \le$ 0.75. The last redshift bin, 0.75$\le z < 1$ is only shown to illustrate our our inability to measure correctly $M^*_{\rm CL}$ at this redshift range.

While it becomes difficult to compare the values of $p_1$ and $p_2$ with different works due to the dependence of the definition of $M^*_{\rm CL}$ on the considered survey \citep{andreon10,ascaso15a}, or other observable used \citep{andreon12,saro15}, we can compare the scatter of the relation. The main scatter found for J-PAS is slightly smaller than this measured with the ALHAMBRA survey  \citep{ascaso15a}. This scatter also becomes comparable to this found by the sample of 52 local ($z<0.1$) clusters and groups ranging a similar mass range \cite{andreon10}.  Other works have found slightly smaller scatter for a sample of clusters ranging a similar range of redshift but five times more massive that in this work. For instance,  \cite{saro15} find $>$ 0.065 dex for very massive $>8\times10^{14}M_{\odot}$ clusters and $>$ 0.087 dex for $>2.5\times10^{14}M_{\odot}$ clusters. The results in this work are encouraging since they show that we can extend previous findings to larger redshift ranges using an optical richness estimator.

In parallel, we have estimated $\sigma_{M_{\rm h} | M^*_{\rm CL}}$, the scatter in the dark matter halo mass at a fixed value of the $M^*_{\rm CL}$. Many authors have noticed the importance of measuring accurately this quantity in order to compute an observational cluster mass function from cluster counts  \citep{rozo09,hilbert10,andreon12}.

\begin{figure}
\centering
\includegraphics[clip,angle=0,width=1.0\hsize]{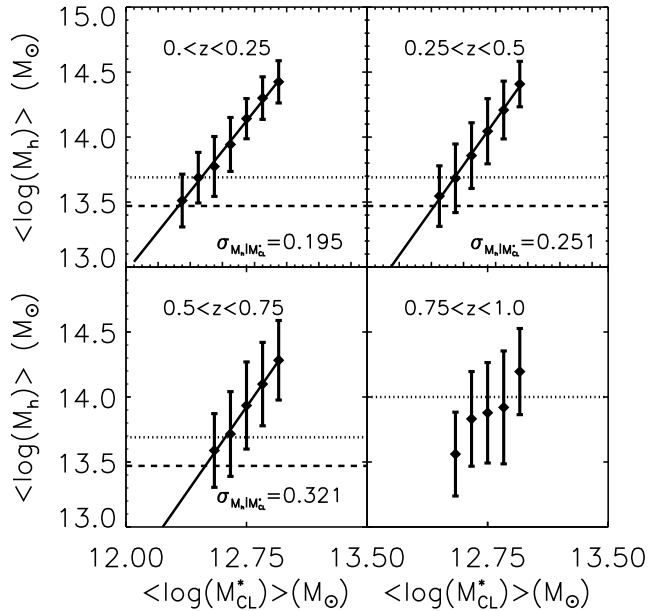} 
\caption{Average halo cluster mass as a function of the average total stellar mass for different redshift bins in logarithmic scale. The solid line displays the linear fits for the two variables. The dotted and dashed lines indicate the mass limit for which the completeness and purity is $>$ 80\% and  $>$70\% respectively for each redshift bin. The average scatter $\sigma_{M_{\rm h} | M^*_{\rm CL}}$, measured as the standard deviation between the two variables is also shown for each redshift bin. The last redshift bin is shown to illustrate the inability to fit the relation at z$>$0.75.}
\label{fig:smassmhalo}
\end{figure}

Following a similar approach as in \cite{ascaso15a}, we have performed 10000 Monte Carlo samplings of the possible halo mass values obtained directly from the simulation (see Fig. \ref{fig:mhalolamda}) to obtain a mean value and scatter for each fixed $M^*_{\rm CL}$. In Fig. \ref{fig:smassmhalo}, we show this calibration for different redshift bins together with the main $\sigma_{M_{\rm h} | M^*_{\rm CL}}$, obtained for each redshift bin.

The mean  $\sigma_{M_{\rm h} | M^*_{\rm CL}}$ value obtained is  $\sim 0.23 \, dex$ down to the mass limit of $M_{\rm h}=5\times10^{13}M_{\odot}$ and  $\sim 0.20 \, dex$ down to the mass limit of $M_{\rm h}=1\times10^{14}M_{\odot}$. This value is somewhat smaller than the values found for the ALHAMBRA survey ($\sigma_{M_{\rm h} | M^*_{\rm CL}} \sim 0.27 \, dex$) down to the same limit and comparable to values found by other authors   for $\sim$5 times higher mass limits.  For instance, \cite{rozo09} found $\sim 0.20 \, dex$ at $N_{200} \sim 40$ (equivalent to $M \sim 2.5 \times 10^{14}M_{ \odot}$) in their calibration between $\log M$ and $N_{200}$, where $M$ is obtained from X-ray and WL proxies and $N_{200}$ is the number of red galaxies lying within $R_{200}$, the radius where the critical density is 200 times the mean density of the universe. Similarly,  \cite{andreon12b} find a scatter of $\sim 0.25  \, dex$ for a sample of 53 local clusters between $M_{200}$ estimated from caustic or velocity dispersion measurements and $N_{200}$. Other authors have recently improved significantly this value for massive clusters. For instance, \cite{saro15}  find $\sim 0.08 \, dex$ at $\lambda \sim 70$ (corresponding to $M \sim 3 \times 10^{14}M_{ \odot}$) between the SZ mass estimate and the redMaPPer \citep{rykoff14} richness estimator, $\lambda$.

Comparing with simulations, similar values have also been found. For instance, \cite{hilbert10} find $\sim 0.22\, dex$ scatter from N-body simulations for the same M and  $N_{200}$ variables down to $M_{\rm h}=5\times10^{13}M_{\odot}$ and  $\sim 0.18\, dex$ scatter from N-body simulations for the same variables down to $M_{\rm h}=1\times10^{14}M_{\odot}$. Likewise, \cite{angulo12} found slightly smaller scatter values ($\sim 0.16 \, dex$) from the Millennium-XXL simulation between the cluster virial mass $M_{200}$ and the optical richness $N_{opt}$ down to $M_{\rm h}\sim 5\times10^{13}M_{\odot}$ for a limited redshift sample (z$<$0.25).

While the simulations might result too simplistic to describe the real data, \cite{ascaso15a} already showed that existing multiple narrow-band surveys, such as ALHAMBRA, are able to decrease substantially the scatter between different optical observables and the dark matter halo mass. Hence, we present results that firmly support the fact that  narrow-band surveys not only allow to detect clusters and groups down to smaller mass limits than broad-band surveys but they also enable to calibrate the observable-halo mass relation with an accuracy comparable to that obtained from broad-band survey for more massive clusters. The imminent start of the survey will confirm this point in the nearby future.

\section{Conclusions}

In this work, we have first characterized the photometric redshift properties of the J-PAS survey in terms of photometric redshift bias, photometric redshift dispersion and rate of catastrophic outliers using an N-body and semi-analytical simulations \citep{merson13} and a posterior modification with \texttt{PhotReal} (\citealt{ascaso15b}, Ben\'itez et al. in prep). We have seen that the mean photometric redshift precision of J-PAS is $\Delta z/(1+z) \sim 0.003$ down to $i\sim$ 23.0, is in agreement with what was expected from previous simulations \citep{benitez09a,benitez14}. Furthermore, the photometric redshift bias is fully consistent with zero down to the same magnitude limit and up to moderate redshift (z$\sim$ 0.7) without performing any preselection  of the survey. The rate of outliers, $\eta_1$, is always lower than 4\% down to $i\sim$ 22.5 and  at least within $0.2\leq z \leq 0.8$.

In addition, we have compared the photometric redshift predictions for J-PAS with similar predictions obtained for the LSST and Euclid, using the same techniques to transform the same original mock catalogues  \citep{ascaso15b}. In this comparison, we conclude that the photometric redshift performance of J-PAS will be outstanding in comparison with other next-generation surveys up to z$\sim$0.7 at least. The photometric redshift dispersion becomes more than 20 times smaller than Euclid+DES or the LSST and more than 10 times smaller than Euclid+DES+LSST surveys together. 

Complementarily, we have also explored the performance of the Bayesian Cluster Finder (BCF) applied to our narrow-band next-generation J-PAS survey. We have demonstrated with realistic simulations  that we will be able to recover groups and masses down to $M_{\rm h}=3\times 10^{13}M_{\odot}$ up to redshifts 0.7 with completeness $>$70\% and purity higher than 80\% and higher masses at $z>0.7$. Restricting completeness to be $>$80\% makes the minimum mass to be  detected to be $M_{\rm h}=5\times 10^{13}M_{\odot}$ up to redshift 0.7. 

We have compared these selection functions with other selection functions coming from different surveys in different wavelengths and we have concluded that J-PAS will reach at least a factor of 2 lower mass threshold than other similar next-generation and present surveys such as the DES and LSST, done with bigger telescopes (4m and 8m respectively). In addition, as the mass function will be sampled to lower mass limits, the absolute total number of detected clusters and groups will be comparable to those detected with the LSST. This is a very important result since the LSST will image more than twice the area of J-PAS. Additionally, since J-PAS will cover a substantial part of the northern sky, whereas the DES and LSST will focus on the southern hemisphere, the J-PAS optical cluster sample will result in an exquisite sample to follow-up clusters detected with eROSITA, for instance.

In addition, we have model the $M^*_{\rm CL} | M_{\rm h}$ relation to a model in order to estimate the relevance of our cluster sample for cosmological purposes. We have considered a log-log normal model with a linear dependence with redshift. The results are compatible with a  non-evolution of this relation with redshift. Also, the main scatter obtained from the limited subsample of clusters down to $M_{\rm h} \sim 5\times 10^{13}M_{\odot}$ and within $0\le z\le 0.75$ is $\sigma_{M^*_{\rm CL} | M_{\rm h}} \sim 0.14\, dex$. These value is comparable to the results presented in other works limited to a local sample \citep{andreon10} or very massive clusters \citep{andreon12,saro15}. This results highlight then the enormous potential of J-PAS for constraining cosmological parameters with galaxy clusters.

Finally, we have also looked into the precision with which we will be able to measure dark matter halo masses by using as observable the total stellar mass of the cluster. The results, based on simulations, suggest that we can recover galaxy clusters halo masses with an average scatter of $\sigma_{M_{\rm h} | M^*_{\rm CL}} \sim 0.23\, dex$ down to $M_{\rm h}\sim 5\times 10^{13}M_{\odot}$.  We note that this quantity becomes comparable to what other work found for samples of  clusters five times more massive both in observations  \citep{rozo09,andreon12b,saro15} and in simulations \citep{hilbert10,angulo12} when restricted to similar mass ranges. Similarly, high accuracies were usually reached with other techniques such as WL \citep{vonderlinden14}, X-rays \citep{rozo14b} or CMB data \citep{planck14}. The impressive calibrations in measuring masses using large number of narrow-bands ($>$50)  photometry provides a new technique that can reinforce the existing ones.

A forthcoming paper (Ascaso et al. in prep) will be devoted to investigate the impact of this selection function on the cosmological parameters, paying particular  attention to the dark energy constraints.

\section*{Acknowledgments}
We thank the referee of this paper for his/her useful and insightful suggestions and comments that helped improving the original manuscript. BA acknowledges financial support for a postdoctoral fellowship from the Observatory of Paris. EC, GLN, CMdO, AM and LS acknowledge funding from PAPESP and CNPq. IO acknowledges support from the European Research Council (ERC) in the form of Advanced Grant, {\sc cosmicism}. We acknowledge support from the Spanish Ministry for Economy and Competitiveness through grants ilink0862 and AYA2013-48623-C02-1-P. BA also thanks the hospitality of the University of Sao Paulo and Observat\'orio Nacional for hosting her for a visit.  BA dedicates this paper to the memory of the J-PAS member Javier Gorosabel.

\end{document}